\documentclass[11pt,fleqn]{iopart}
\usepackage[utf8]{inputenc}
\usepackage[T1]{fontenc}
\expandafter\let\csname equation*\endcsname=\relax
\expandafter\let\csname endequation*\endcsname=\relax
\usepackage{amsmath,amssymb,amsfonts,amsthm}
\usepackage{bbold}
\usepackage{enumerate}
\usepackage{MnSymbol}
\newcommand{\ket}[1]{\,\rvert#1\rangle}
\newcommand{\bra}[1]{\,\langle #1\rvert}
\newcommand{\braket}[2]{\langle\,#1\,\rvert\,#2\,\rangle}
\newcommand{\Braket}[3]{\bra{#1}\,#2\,\ket{#3}}
\numberwithin{equation}{section}
\begin{document}
\title[Quantum spin chains with fractional revival]{Quantum spin chains with fractional revival}
\author{Vincent X. Genest}
\ead{vincent.genest@umontreal.ca}
\address{Centre de recherches math\'ematiques, Universit\'e de Montr\'eal, Montr\'eal (QC), Canada}
\author{Luc Vinet}
\ead{luc.vinet@umontreal.ca}
\address{Centre de recherches math\'ematiques, Universit\'e de Montr\'eal, Montr\'eal (QC), Canada}
\author{Alexei Zhedanov}
\ead{zhedanov@yahoo.com}
\address{Donetsk Institute for Physics and Technology, Donetsk 83114, Ukraine}
\begin{abstract}
A systematic study of fractional revival at two sites in $XX$ quantum spin chains is presented and analytic models with this phenomenon are exhibited. The generic models have two essential parameters and a revival time that does not depend on the length of the chain. They are obtained by combining two basic ways of realizing fractional revival in a spin chain each bringing one parameter. The first proceeds through isospectral deformations of spin chains with perfect state transfer. The second arises from the recurrence coefficients of the para-Krawtchouk polynomials with a bi-lattice orthogonality grid. It corresponds to an analytic model previously identified that can possess perfect state transfer in addition to fractional revival.
\end{abstract}
\section{Introduction}
Spin chains with engineered couplings have proved attractive for the purpose of designing devices to achieve quantum information tasks such as quantum state transfer or entanglement generation \cite{2007_Bose_ContempPhys_48_13, 2010_Kay_IntJQtmInf_8_641, 2014_Nikilopoulos&Jex}. One reason for the interest is that the internal dynamics of the chain takes care of the processes with a minimum of external intervention required. In this perspective, a desired feature of such chains is that they exhibit quantum revival. For perfect state transfer (PST), one wishes to have, for example, a one-excitation state localized at the beginning of the chain evolve with unit probability, after some time $T$, into the state with the excitation localized at the end of the chain. Such a relocalization of the wave packet is what is referred to as revival \cite{2004_Robinett_PhysReports_392_1,2001_Berry&Marzoli&Schleich_PhysWorld_14_39}. Fractional revival occurs when a number of smaller packets seen as little clones of the original one form at certain sites and show local periodicities \cite{2004_Robinett_PhysReports_392_1, 1997_Aronstein&Stroud_PhysRevA_55_4526}. Its realization in a spin chain would also allow to transport information with high efficiency via one of the clones. Moreover, in a balanced case where there is equal probability of finding clones at the beginning and at the end of a chain, fractional revival would provide a mechanism to generate entangled states. It is hence of relevance to determine if FR is feasible in spin chains, and if so, in which models. Some studies have established the fact that this effect can indeed be observed in spin chains  \cite{2010_Kay_IntJQtmInf_8_641, 2007_Chen&Song&Sun_PhysRevA_75_012113, 2015_Banchi&Compagno&Bose_PhysRevA_91_052323, GVZ-2015}. The present paper offers a systematic analysis of the circumstances under which fractional revival at two sites in quantum spin chains of the $XX$ type will be possible. The main question is to determine the Hamiltonians $H$ that will have the fractional revival property. Like for PST, the conditions for FR are expressed through requirements on the one-excitation spectrum of $H$. One then deals with an inverse spectral problem that can be solved with the assistance of orthogonal polynomial theory. In the full revival or PST case where this analysis has been carried out in detail (see \cite{2012_Vinet&Zhedanov_PhysRevA_85_012323} for instance), one sees that a necessary condition is that the couplings and Zeeman terms must form a three-diagonal matrix that is mirror-symmetric. Furthermore, models based on special orthogonal polynomials have been found where PST can be exhibited in an exact fashion. These analytic models are quite useful; in fact, the simplest one \cite{2004_Albanese&Christandl&Datta&Ekert_PhysRevLett_93_230502} was employed to perform an experimental quantum simulation with Nuclear Magnetic Resonance techniques of mirror inversion in a spin chain \cite{2014_Rao&Mahesh&Kumar_PhysRevA_90_012306}. We shall here also provide analytic models with FR.

The outline of the paper is as follows. In Section 2, we present the Hamiltonians $H$ for the class of $XX$ spin chains that will be considered. The  reader is reminded that their one-excitation restrictions $J$ are given by Jacobi matrices and the orthogonal polynomials associated to the diagonalization of those $J$'s are described. In Section 3, we review the elements of the characterization of $XX$ spin chains with PST that will be essential in our fractional revival study. In particular, the necessary and sufficient conditions for PST in terms of the spectrum of $J$, the mirror symmetry and the properties of the associated orthogonal polynomials will be recalled. 

In Section 4, we undertake a systematic analysis of the conditions under which fractional revival can occur at two sites. Up to a global phase, the revived states will be parametrized in terms of two reals amplitudes $\sin \theta$ and $\cos \theta$ and of a relative phase $\psi$; the case $\theta=0$ will correspond to the PST situation. It will be shown that in general, for FR to occur, the one-excitation spectrum of the Hamiltonian must take the form of a bi-lattice, i.e. the spectral points need to be the union of two uniform lattices translated one with respect to the other by a parameter $\delta$ depending on $\theta$ and $\psi$. Two special cases, namely $\psi=0$ and $\psi=\pi/2$, will be the object of the subsequent two sections. In the first case the spectrum condition is the PST one and in the second case, it is rather the mirror symmetry that is preserved. These two cases will provide the ingredients of a two-step process for obtaining the most general $XX$ chains with FR at two sites.

In Section 5, it will be shown how spin chains with FR can be obtained from isospectral deformations of spin chains with PST. This has the relative phase $\psi=0$. Analytic models with FR will thus be obtained from analytic models with PST. It will be observed that only the central parameters of the chain will need to be modified so as to make FR happen. Mirror-symmetry will be seen to be replaced by a more complicated inversion. The orthogonal polynomials corresponding to the deformed Jacobi matrix will be shown to have a simple expression in terms of the unperturbed orthogonal polynomials associated to the parent PST Hamiltonian. Their knowledge will be relevant to the construction, at the end of Section 6, of the FR Hamiltonian with arbitrary parameters $\theta$ and $\psi$.

Section 6 will deal with the special case when the relative phase is $\pi/2$, that is when one amplitude is real and the other purely imaginary. It will be seen that $J$ is mirror symmetric in this situation with its spectrum a bi-lattice. As shall be explained, an exact solution to the corresponding inverse spectral problem turns out to be already available and is provided by the recurrence coefficients of the para-Krawtchouk polynomials introduced in \cite{2012_Vinet&Zhedanov_JPhysA_45_265304}. It is remarkable that these somewhat exotic functions naturally arise in the FR analysis. We shall demonstrate that for certain values of the parameters, the models thus engineered possess both FR and PST. We shall finally come to the determination of the generic Hamiltonian with FR and show that is is obtained from the recurrence coefficients of para-Krawtchouk polynomials perturbed in the way described in Section 5; in other words it is constructed by performing an isospectral deformation of the Jacobi matrix of the para-Krawtchouk polynomials. The two arbitrary parameters are related to the the bi-lattice and to the deformation parameter. We shall summarize the outcome of the analysis and offer final remarks to conclude.
\section{XX quantum spin chain models with non-uniform nearest-neighbor interactions}
We shall consider $XX$ spin chains with $N+1$ sites and nearest-neighbor interactions that are governed by Hamiltonians $H$ of the form
\begin{align}
\label{H}
H=\frac{1}{2}\sum_{\ell=0}^{N-1}J_{\ell+1}(\sigma_{\ell}^{x}\sigma_{\ell+1}^{x}+\sigma_{\ell}^{y}\sigma_{\ell+1}^{y})+\frac{1}{2}\sum_{\ell=0}^{N}B_{\ell}(\sigma_{\ell}^{z}+1)
\end{align}
that act on $(\mathbb{C}^2)^{\otimes N+1}$. The $J_{\ell}$ are the coupling constants between the sites $\ell-1$ and $\ell$ and the $B_{\ell}$ are the strengths of the magnetic fields at the sites $\ell$, where $\ell=0,1,\ldots, N$. The index $\ell$ on the Pauli matrices $\sigma^{x}$, $\sigma^{y}$, $\sigma^{z}$ indicates on which of the $(N+1)$ $\mathbb{C}^2$ factor these matrices act. If $\ket{\uparrow}$ and $\ket{\downarrow}$ denote the eigenstates of $\sigma^{z}$ with eigenvalues $+1$ and $-1$ respectively, $\sigma^{x}$ and $\sigma^{y}$ are known to act as follows in that basis: $\sigma^{x}\ket{\uparrow}=\ket{\downarrow}$, $\sigma^{x}\ket{\downarrow}=\ket{\uparrow}$; $\sigma^{y}\ket{\uparrow}=-i\ket{\downarrow}$, $\sigma^{y}\ket{\downarrow}=i\ket{\uparrow}$. The Hamiltonians $H$ are invariant under rotations about the $z$-axis, i.e.
\begin{align}
[H,\frac{1}{2}\sum_{\ell=0}^{N}(\sigma_{\ell}^{z}+1)]=0,
\end{align}
and as a consequence, the eigenstates of $H$ split in subspaces labeled by the number of spins over the chain that are in the state $\ket{\uparrow}$, which is a conserved quantity. In the following we shall focus on the subspace with one excitation and we will denote by $J$ the restriction of $H$ to that subspace, equivalent to $\mathbb{C}^{N+1}$. A natural orthonormal basis for the states with one excitation is given by the vectors 
\begin{align}
\ket{\ell}=(0,0,\ldots, 1,\ldots,0)^{\top},\qquad \ell=0,1,\ldots, N,
\end{align}
where the only ``1'' in the $\ell$\textsuperscript{th} entry corresponds to the only state $\ket{\uparrow}$ being at the site $\ell$. The action of $J$ on these basis vectors is directly obtained from \eqref{H} and given by
\begin{align}
J\ket{\ell}=J_{\ell+1}\ket{\ell+1}+B_{\ell}\ket{\ell}+J_{\ell}\ket{\ell-1},
\end{align}
where it is assumed that $J_0=J_{N+1}=0$. The restricted Hamiltonian $J$ thus takes the form of a $(N+1)\times (N+1)$ three-diagonal Jacobi matrix
\begin{align}
\label{Tridiag}
J=
 \begin{pmatrix}
B_0	&	J_1	&		&	&
\\
J_1	&	B_1	&	J_2	&	&
\\
	&	J_2	&	B_2	& \ddots &
\\
	& 		& \ddots		& 	\ddots & J_{N}
\\
 & & &J_{N}&B_{N}
 \end{pmatrix}.
\end{align}
It is known that Jacobi matrices are diagonalized by orthogonal polynomials. Let us record, in this connection, results that will prove useful in our study. Let $\ket{\lambda}$ be the eigenvectors of $J$:
\begin{align}
\label{Eigen-General}
J\ket{\lambda}=\lambda\ket{\lambda}.
\end{align}
Since $J$ is Hermitian the eigenvalues $\lambda$ are real. We shall assume that they are non-degenerate, i.e. that they take $N+1$ different values $\lambda_{s}$ for $s=0,1,\ldots,N$. This is the case when the entries $J_{\ell}$ are positive. We will take the eigenvalues to be ordered 
\begin{align}
\lambda_0<\lambda_1<\cdots <\lambda_N.
\end{align}
Consider the expansion of $\ket{\lambda_{s}}$ in terms of the basis vectors $\ket{\ell}$,
\begin{align}
\label{Expansion}
\ket{\lambda_{s}}=\sum_{\ell=0}^{N}W_{s\ell}\,\ket{\ell}
\end{align}
and write the elements $W_{s\ell}$ of the transition matrix in the form
\begin{align*}
W_{s\ell}=W_{s0}\,\chi_{\ell}(\lambda_{s})\equiv \sqrt{\omega_{s}}\chi_{\ell}(\lambda_{s}).
\end{align*}
It follows from \eqref{Eigen-General} that $\chi_{\ell}(\lambda_{s})$ are polynomials satisfying the three-term recurrence relation
\begin{align}
J_{\ell+1}\,\chi_{\ell+1}(\lambda_{s})+B_{\ell}\,\chi_{\ell}(\lambda_{s})+J_{\ell}\,\chi_{\ell-1}(\lambda_{s})=\lambda\, \chi_{\ell}(\lambda_{s}),
\end{align}
with initial condition
\begin{align}
\chi_{0}=1,\qquad \chi_{-1}=0.
\end{align}
Since both the eigenbasis $\{\ket{\lambda_{s}}\}_{s=0}^{N}$ and the basis $\{\ket{\ell}\}_{\ell=0}^{N}$ are orthonormal and all the coefficients are real, the matrix $(W)_{s\ell}$ is orthogonal and the condition $\braket{\ell}{\ell'}=\delta_{\ell\ell'}$ implies that the polynomials $\chi_{\ell}(\lambda)$ are orthogonal on the finite set of spectral points $\lambda_{s}$:
\begin{align}
\label{Ortho}
\sum_{s=0}^{N}w_{s}\chi_{m}(\lambda_{s})\chi_{n}(\lambda_{s})=\delta_{mn}.
\end{align}
Note that the weights $w_{s}$ are normalized, that is
\begin{align}
\label{Norm-Cond}
\sum_{s=0}^{N}w_{s}=1,
\end{align}
since $\chi_{0}(\lambda_{s})=1$. Owing to the orthogonality of $W$, in addition to \eqref{Expansion}, we also have
\begin{align}
\label{Expansion-2}
\ket{\ell}=\sum_{s=0}^{N}W_{s\ell}\,\ket{\lambda_{s}}=\sum_{s=0}^{N}\sqrt{w_{s}}\,\chi_{n}(\lambda_{s})\ket{\lambda_{s}}.
\end{align}
It is sometimes useful to use the monic polynomials 
\begin{align}
\label{14}
P_{\ell}(\lambda)=\sqrt{h_{\ell}}\,\chi_{\ell}(\lambda),\qquad \sqrt{h_{\ell}}=J_1J_2\cdots J_{\ell},
\end{align}
whose leading coefficient is 1: $P_{\ell}(\lambda)=\lambda^{\ell}+\cdots $. The spectrum of $J$ is encoded in the characteristic polynomial 
\begin{align}
P_{N+1}(\lambda)=(\lambda-\lambda_0)(\lambda-\lambda_1)\cdots (\lambda-\lambda_{N}).
\end{align}
The following formula from the standard theory of orthogonal polynomials  gives the weights $w_{s}$ in terms of $P_{N}(\lambda)$ and $P_{N+1}(\lambda)$:
\begin{align}
\label{W-1}
w_{s}=\frac{h_{N}}{P_{N}(\lambda_{s})P_{N+1}'(\lambda_{s})},\qquad s=0,1,\ldots, N,
\end{align}
with $P_{N+1}'(\lambda)$ denoting the derivative of $P_{N+1}(\lambda)$ with respect to $\lambda$ \cite{Chihara-2011}.
\section{A review of perfect state transfer in a $XX$ spin chain}
Perfect state transfer is achieved along a chain if there is a time $T$ for which
\begin{align}
\label{PST-Cond}
e^{-iTJ}\ket{0}=e^{i\phi}\ket{N},
\end{align}
where $\phi$ is a real number. In this case the initial state with one spin up at the zeroth site will be found with unit probability after time $T$ in the state with one spin up at the site $N$. One then says that the state $\ket{\uparrow}$ has been perfectly transferred from one end of the chain to the other. The requirements for PST have been determined from studying the implications of \eqref{PST-Cond}. Since this will be the backdrop for our discussion of fractional revival, it is pertinent to summarize the PST analysis here. Using the expansion \eqref{Expansion-2} on both sides of \eqref{PST-Cond} gives
\begin{align}
\label{Condition-Chi}
e^{-i\phi}e^{-iT\lambda_{s}}=\chi_{N}(\lambda_{s}),\qquad s=0,1,\ldots, N.
\end{align}
Since $\chi_{N}(\lambda_{s})$ is real,  this implies that 
\begin{align}
\label{Condition-Chi-2}
\chi_{N}(\lambda_{s})=\pm 1.
\end{align}
Now because the zeros of $\chi_{N}(\lambda)$ must lie  between those of $\chi_{N+1}(\lambda)$ which are located at the $\lambda_{s}$, we must conclude that $\chi_{N}(\lambda_{s})$ will alternate between $+1$ and $-1$ \cite{Chihara-2011}. Given that the eigenvalues are ordered, the condition that the weights, as given by \eqref{W-1}, must be positive leads finally to 
\begin{align}
\label{Eigen-Chi}
\chi_{N}(\lambda_{s})=(-1)^{N+s},\qquad s=0,1,\ldots, N.
\end{align}
In view of \eqref{Eigen-Chi}, we see that \eqref{Condition-Chi} imposes the following requirement on the spectrum of $J$:
\begin{align}
\label{Eigen-Lambda}
e^{-iT\lambda_{s}}=e^{i\phi}(-1)^{N+s},\qquad s=0,1,\ldots,N,
\end{align}
which is tantamount to the condition that the successive eigenvalues are such that
\begin{align}
\label{Eigen-Cond}
\lambda_{s+1}-\lambda_{s}=\frac{\pi}{T} M_{s},
\end{align}
with $M_{s}$ an arbitrary positive odd integer. We have thus with \eqref{Eigen-Chi} and \eqref{Eigen-Lambda}, the necessary and sufficient conditions for PST to occur. It is seen from \eqref{W-1} that the necessary condition \eqref{Eigen-Chi} is equivalent to requiring that the polynomials associated to the Jacobi matrix $J$ be orthogonal with respect to the weights
\begin{align}
\label{Weight-Explicit}
w_{s}=\frac{\sqrt{h_{N}}\, (-1)^{N+s}}{P_{N+1}'(\lambda_s)},\quad s=0,1,\ldots, N.
\end{align}
Remarkably, the weights given by \eqref{Weight-Explicit} have the general property \cite{Tsujimoto_2015}
\begin{align}
\label{Gen-Prop}
\sum_{s}w_{2s}=\sum_{s}w_{2s+1}=\frac{1}{2}.
\end{align}
In turn, \eqref{Eigen-Chi} or \eqref{Eigen-Cond} can be shown to hold if and only if the matrix $J$ is mirror symmetric with respect to the anti-diagonal \cite{2012_Vinet&Zhedanov_PhysRevA_85_012323}, that is if and only if 
\begin{align}
\label{Per-Sym}
RJR=J,
\end{align}
with
\begin{align}
R=
 \begin{pmatrix}
  &&&1\\
  &&1&\\
  &\udots&&
  \\
  1 &&&
 \end{pmatrix}.
\end{align}
In terms of the couplings and magnetic field strengths, one sees from \eqref{Tridiag} that this symmetry amounts to the relations
\begin{align}
\label{Mir-Sym}
J_{n}=J_{N+1-n},\qquad B_{n}=B_{N-n}.
\end{align}
Matrices with that property are referred to as persymmetric matrices. A simple and direct proof that \eqref{PST-Cond} implies \eqref{Per-Sym} can be found in \cite{2010_Kay_IntJQtmInf_8_641}. 

Let us now observe that this required symmetry will lead not only to PST but to a complete mirror inversion of the register at time $T$ if the spectral condition \eqref{Eigen-Lambda} is satisfied. First note that $R$ is an involution, that is $R^2=\mathbb{1}$. Because $J$ is reflection-invariant, its eigenstates can be taken to also be eigenstates of $R$ and since we have assumed that the spectrum of $J$ is not degenerate, each eigenstate must hence be of definite parity $\epsilon_{s}$, equal to either $+1$ or $-1$. We hence have
\begin{align}
\label{abv-5}
R\ket{\lambda_{s}}=\epsilon_{s}\ket{\lambda_{s}}.
\end{align}
From \eqref{Expansion}, we have
\begin{align}
R\ket{\lambda_{s}}=\sum_{\ell=0}^{N}\sqrt{w_{s}}\,\chi_{\ell}(\lambda_{s})\ket{N-\ell}
=\sum_{\ell=0}^{N}\sqrt{w_{\ell}}\,\chi_{N-\ell}(\lambda_{s})\ket{\ell},
\end{align}
but given \eqref{abv-5}, using \eqref{Expansion} again we find that
\begin{align}
\label{Chi-Prop}
\chi_{N-\ell}(\lambda_{s})=(-1)^{N+s}\chi_{\ell}(\lambda_{s}),
\end{align}
where we have determined $\epsilon_{s}$ by setting $\ell=0$ and using \eqref{Eigen-Chi}. As a result, we see that 
\begin{multline}
\label{Calculation}
\Braket{k}{e^{-iTJ}}{\ell}=\sum_{\ell=0}^{N} e^{-iT\lambda_{s}}\,w_{s}\,\chi_{\ell}(\lambda_{s})\chi_{k}(\lambda_{s})
=e^{i\phi}\sum_{s=0}^{N}(-1)^{N+s}w_{s}\,\chi_{\ell}(\lambda_{s})\chi_k(\lambda_{s})
\\
=e^{i\phi}\sum_{s=0}^{N}w_{s}\,\chi_{N-\ell}(\lambda_{s})\chi_{k}(\lambda_{s})=e^{i\phi}\delta_{k,N-\ell},
\end{multline}
with the successive help of \eqref{Eigen-Lambda}, \eqref{Chi-Prop} and \eqref{Ortho}. In other words, there is probability 1 of finding at the site $N-n$, after time $T$, the spin up initially at site $n$. In matrix form, we have shown that \eqref{Eigen-Chi} or \eqref{Eigen-Cond} and \eqref{Eigen-Lambda} have for consequence that
\begin{align}
e^{-iTJ}=e^{i\phi}R.
\end{align}
Now the main issue is to determine the Hamiltonians $H$ of type \eqref{H} that have the PST property. This is an inverse spectral problem since we start from conditions on the eigenvalues. As it turns out, this specific problem has been well studied \cite{2004_Gladwell}. The outcome is that once a spectrum satisfying \eqref{Eigen-Lambda} or \eqref{Eigen-Cond} is given as input, the Hamiltonian with the desired properties is uniquely determined. This stems from the fact that the weights $w_{s}$ entailing mirror-symmetry are uniquely prescribed by \eqref{Weight-Explicit} and that the corresponding orthogonal polynomials can be unambiguously constructed. Their recurrence coefficients henceforth provide the couplings $J_{\ell}$ and the local magnetic fields $B_{\ell}$ of a Hamiltonian $H$ with PST. One algorithm for obtaining the orthogonal polynomials is described in \cite{2012_Vinet&Zhedanov_PhysRevA_85_012323}. By considering natural types of spectrum, analytic models of spin chains with PST have thus been obtained and associated to various families of orthogonal polynomials \cite{2010_Koekoek_&Lesky&Swarttouw}: Krawtchouk and Dual Hahn \cite{2004_Albanese&Christandl&Datta&Ekert_PhysRevLett_93_230502}, Dual $-1$ Hahn \cite{2005_Shi&Li&Song&Sun_PhysRevA_71_032309,2011_Stoilova&VanderJeugt_SIGMA_7_33, 2012_Vinet&Zhedanov_JPhysConfSer_343_012125}, $q$-Krawtchouk \cite{2010_Jafarov&VanderJeugt_JPhysA_43_405301} and $q$-Racah \cite{2012_Vinet&Zhedanov_PhysRevA_85_012323}. Of particular interest for what follows are models connected to the so-called para-Krawtchouk polynomials \cite{2012_Vinet&Zhedanov_JPhysA_45_265304}. The paradigm example of analytic models, investigated in \cite{2004_Albanese&Christandl&Datta&Ekert_PhysRevLett_93_230502}, is obtained by considering the linear spectrum
\begin{align}
\lambda_{s}=\frac{\pi}{T}\Big(s-N/2\Big),\qquad s=0,1,\ldots, N,
\end{align}
for which \eqref{Weight-Explicit} yields the binomial distribution
\begin{align}
w_{s}=\frac{N!}{s!(N-s)!}\left(\frac{1}{2}\right)^{N}.
\end{align}
The associated polynomials are known to be the symmetric Krawtchouk polynomials with recurrence coefficients
\begin{align}
\label{Recu-Krawtchouk}
B_{\ell}=0,\quad J_{\ell}^2=\frac{\pi^2}{T^2}\frac{\ell(N+1-\ell)}{4}.
\end{align}
The mirror symmetry \eqref{Mir-Sym} is manifest. This is one instance where the one-excitation spectrum dynamics is exactly solvable. The general transition amplitude from state $\ket{\ell}$ to state $\ket{k}$ in time $t$ has been calculated in \cite{2010_Chakrabarti&VanderJeugt_JPhysA_43_085302} to be
\begin{multline}
\label{Kraw-Amp}
\Braket{k}{e^{-itJ}}{\ell}=\left(\frac{1}{2}\right)^{N}\sqrt{\binom{N}{k}\binom{N}{\ell}}\;(1-e^{-i\frac{T}{\pi}t})^{k+\ell}
\\
\times \,(1+e^{-i\frac{T}{\pi}t})^{N-k-\ell}\;
{}_2F_{1}\left(\genfrac{}{}{0pt}{}{-k,-\ell}{-N},\;\frac{-4e^{-i\frac{T}{\pi}t}}{(1-e^{-i\frac{T}{\pi}t})^2}\right),
\end{multline}
where ${}_2F_{1}$ is the classical hypergeometric series \cite{2001_Andrews&Askey&Roy}.

It must be stressed finally that a myriad of analytic models can be generated from those associated to known orthogonal polynomials by a procedure known as ``spectral surgery'' \cite{2012_Vinet&Zhedanov_PhysRevA_85_012323}. Indeed, it has been shown that the first or last eigenvalues ($\lambda_0$ or $\lambda_N$) or a pair of neighboring spectral points ($\lambda_i,\lambda_{i+1}$) can be removed without affecting the PST properties \cite{2012_Vinet&Zhedanov_PhysRevA_85_012323}. If $P_{\ell}(\lambda)$ are the monic polynomials associated to the original chain with Hamiltonian $J$, the removal of one level, say $\lambda_i$, will give a new Jacobi matrix $J'$ with entries given by
\begin{align}
\label{C-1}
(J_{\ell}')^2=\left(\frac{A_{\ell}}{A_{\ell-1}}\right)\,J_{\ell}^2 ,\quad B_{\ell}'=B_{\ell+1}+A_{\ell+1}-A_{\ell},
\end{align}
where
\begin{align}
A_{\ell}=\frac{P_{\ell+1}(\lambda_i)}{P_{\ell}(\lambda_i)}.
\end{align}
The monic polynomials corresponding to $J'$ are
\begin{align}
\label{C-2}
P_{\ell}'(x)=\frac{P_{\ell+1}(x)-A_{\ell}P_{\ell}(x)}{x-x_i}.
\end{align}
Formulas \eqref{C-1} and \eqref{C-2} can be applied iteratively to obtain new analytic models from known ones. The removal of pairs of neighboring spectral points is required in the bulk to ensure the positivity of the weights.
\section{Spectral conditions for fractional revival}
We are now ready to examine the requirements for fractional revival in a spin chain of type $XX$. We shall limit ourselves to revivals occurring at only two sites chosen to be $\ell=0$ and $\ell=N$, the two ends of the chain. We shall here analyze systematically the conditions that will ensure that for a time $T$
\begin{align}
\label{FR-C1}
e^{-iTJ}\ket{0}=\alpha \ket{0}+\beta \ket{N},
\end{align}
with $|\alpha|^2+|\beta|^2=1$. When this is so, the spin up at site $\ell=0$ at $t=0$ is revived at the sites $\ell=0$ and $\ell=N$ after time $T$ with amplitude $\alpha$ and $\beta$, respectively. Let us write 
\begin{align}
\alpha=e^{i\phi}\sin 2\theta,\qquad \beta=e^{i(\phi+\psi)}\cos 2\theta,
\end{align}
with $e^{i\phi}$ and $e^{i\psi}$ the global and relative phase factors of the two complex numbers $\alpha$ and $\beta$. Taking $-\frac{\pi}{4}\leq \theta \leq \frac{\pi}{4}$ and $0\leq \psi <\pi$ with $\phi\in \mathbb{R}$ covers all possible amplitudes. The use of \eqref{Expansion-2} leads to the relation
\begin{align}
\label{Cond-Eigenvalues-2}
e^{-iT\lambda_{s}}=e^{i\phi}\left[\sin 2\theta +e^{i\psi}\cos 2\theta\,\chi_{N}(\lambda_s)\right],\qquad s=0,1,\ldots, N.
\end{align}
This implies that $\sin 2\theta +e^{i\psi}\cos 2\theta\,\chi_{N}(\lambda_s)$ has modulus 1, that is
\begin{align}
\label{Condition-Chi-3}
\chi_{N}^2(\lambda_s)+2\tan 2\theta \cos\psi-1=0.
\end{align}
Obviously when $\theta=0$, we recover condition \eqref{Condition-Chi-2}. Equation \eqref{Condition-Chi-3} indicates that $\chi_{N}(\lambda_{s})$ will take one of two values, $\gamma$ and $-\frac{1}{\gamma}$, with $\gamma$ satisfying 
\begin{align}
\label{Cond-Gamma}
\gamma-\frac{1}{\gamma}=-2 \tan 2\theta \cos \psi.
\end{align}
By a continuity argument as $\theta\rightarrow 0$, upon comparing with \eqref{Eigen-Chi} and assuming that $\gamma$ is the positive root (there must be one because of the interlacing property of the zero of orthogonal polynomials), one concludes that, for $N$ odd
\begin{subequations}
\label{Cond-Gamma-2}
\begin{align}
\chi_{N}(\lambda_{2s})=-\frac{1}{\gamma},\qquad \chi_{N}(\lambda_{2s+1})=\gamma,
\end{align}
for $N$ even
\begin{align}
\chi_{N}(\lambda_{2s})=\gamma,\qquad \chi_{N}(\lambda_{2s+1})=-\frac{1}{\gamma},
\end{align}
\end{subequations}
for all $s\in \{0,\ldots, N\}$. Assume for now that the relative phase factor $e^{i \psi}$ is generic. Once \eqref{Cond-Gamma} and \eqref{Cond-Gamma-2} are satisfied, it must still be ensured that \eqref{Cond-Eigenvalues-2} is obeyed. When $N$ is odd, this amounts to the conditions
\begin{subequations}
\label{abcd}
\begin{align}
\label{a}
&\cos(T\lambda_{2s}+\phi)=\sin 2\theta-\frac{1}{\gamma} \cos 2\theta \cos \psi,
\quad
\sin(T\lambda_{2s}+\phi)=\frac{1}{\gamma}\cos 2\theta \sin \psi,
\\
\label{c}
&\cos(T\lambda_{2s+1}+\phi)=\sin 2\theta+\gamma \cos 2\theta \cos \psi,
\quad
\sin(T\lambda_{2s+1}+\phi)=-\gamma \cos 2\theta \sin \psi.
\end{align}
\end{subequations}
Define $\xi, \eta\in [0,2\pi]$ by setting the right-hand sides of equations \eqref{a}, \eqref{c} to be respectively $\cos \xi$, $\sin \xi$, $\cos \eta$ and $\sin \eta$. This implies
\begin{align}
\label{sets}
\{T\lambda_{2s}+\phi\}\subseteq \{\xi, \xi\pm 2\pi, \xi \pm 4 \pi, \ldots \},
\quad 
\{T \lambda_{2s+1}+\phi\}\subseteq\{\eta,\eta\pm 2\pi,\eta\pm 4\pi,\ldots \}.
\end{align}
When $N$ is even, it is seen that the roles of $\xi$ and $\eta$ are interchanged. On thus observes that the spectrum of $J$ must be the following bi-lattices:
\begin{align}
\label{Bi-Lattice}
\frac{T\lambda_{s}+\phi}{\pi}=\frac{\mu}{\pi}+s+\frac{1}{2}(\delta-1)(1-(-1)^{s}),
\end{align}
for $s=0,1,\ldots, N$ with
\begin{align}
\label{mu}
\mu=
\begin{cases}
\xi & \text{for $N$ odd}
\\
\eta & \text{for $N$ even}
\end{cases},
\end{align}
and 
\begin{align}
\delta=
\begin{cases}
2+\frac{1}{\pi}(\eta-\xi) & \text{for $N$ odd}
\\
\frac{1}{\pi}(\xi-\eta) & \text{for $N$ even}
\end{cases}.
\end{align}
We have used the latitude in picking the initial point in the sets \eqref{sets} and those obtained by the exchange $\xi\leftrightarrow \eta$ for $N$ even, to ensure that $\delta$ is positive. Let us point out that the spectra can in fact also be subsets of the bi-lattices \eqref{Bi-Lattice} after appropriate surgery. 

There are two special cases for the phase $\psi$ that are of particular interest and that will be the object of the next sections. Each preserve one of the two necessary and sufficient conditions \eqref{Eigen-Chi} and \eqref{Eigen-Lambda} for PST. These two distinguished cases are $\psi=0$ and $\psi=\pi/2$. Let us make initial observations about what happens for those values.
\begin{enumerate}[i.]
\item $\psi=\frac{\pi}{2}$
\end{enumerate}
It is readily seen from \eqref{Cond-Gamma} and \eqref{Cond-Gamma-2} that $\gamma=1$ and that $\chi_{N}(\lambda_{s})=(-1)^{N+s}$. Hence mirror-symmetry is maintained in this case. This is the case considered in \cite{2015_Banchi&Compagno&Bose_PhysRevA_91_052323}. Equations \eqref{abcd} give that
\begin{align}
\xi=-\eta=\frac{\pi}{2}-2\theta.
\end{align}
The spectral points thus form bi-lattices of the form \eqref{Bi-Lattice} with 
\begin{align}
\label{delta}
\delta=1\pm \frac{4\theta}{\pi}.
\end{align}
The upper/lower sign in the above equation corresponds to $N$ being odd/even. Note that $\delta \in [0,2]$ for $-\frac{\pi}{4}\leq \theta \leq \frac{\pi}{4}$; when $\theta=0$, it is the PST situation, $\delta=1$ and \eqref{Bi-Lattice} becomes the linear spectrum of the Krawtchouk polynomials with $\phi=\frac{\pi}{2}(N\pm 1)$. As shall be explained in Section 6, the model corresponding to the spectral conditions for $\psi=\frac{\pi}{2}$ is analytic and can exhibit both FR and PST.
\begin{enumerate}[ii.]
\item $\psi=0$
\end{enumerate}
In this case \eqref{Cond-Gamma} gives for $\gamma$
\begin{align}
\label{abv-2}
\gamma=\cot \left(\frac{\pi}{4}-\theta\right),
\end{align}
and in view of \eqref{Cond-Gamma-2} mirror symmetry is broken. However using \eqref{abv-2} and \eqref{Cond-Gamma-2} it is immediate to check that \eqref{Cond-Eigenvalues-2} becomes
\begin{align}
e^{-iT\lambda_{s}}=e^{i\phi}(-1)^{N+s},
\end{align}
which is the spectral condition \eqref{Eigen-Lambda} for PST. So when $\psi=0$, PST is absent for lack of mirror symmetry but the spectrum remains unchanged. This isospectral situation is discussed next.
\section{Isospectral deformations of chains with perfect state transfer}
We shall now describe the spin chains with fractional revival that can be obtained by isospectral deformations of chains with PST \cite{GVZ-2015}. This picture arises when the relative phase $\psi$ is nil. It should be stressed from the outset that the procedure will generate analytic models with FR when it is applied to spin chains for which PST can be exactly demonstrated. In this section, for the sake of clarity, we shall denote by $\widetilde{J}$ the one-excitation Hamiltonians of spin chains with FR and by $J$ those of spin chains with PST. When $\psi=0$, the FR condition \eqref{FR-C1} reads
\begin{align}
\label{FR-2}
e^{-iT\widetilde{J}}\ket{0}=e^{i\phi}\left[\sin 2\theta\, \ket{0}+\cos 2\theta\, \ket{N}\right]
\end{align}
and \eqref{Cond-Eigenvalues-2} becomes
\begin{align}
\label{ebv-3}
e^{-i\phi}e^{-iT\lambda_{s}}=\sin 2\theta+\cos 2\theta\,\chi_{N}(\lambda_{s}).
\end{align}
We observed that because the right-hand side of \eqref{ebv-3} is real, the condition on the spectrum of $\widetilde{J}$ is the same as for PST, that is \eqref{Eigen-Lambda}. For one such spectrum, still assumed to be non-degenerate, it must be possible to relate by a conjugation the Jacobi matrix with PST to the one with FR. Recall that the PST matrix $J$ can be uniquely constructed from the data. There is thus an orthogonal matrix $U$ such that
\begin{align}
\widetilde{J}=U J U^{\top}.
\end{align}
It then follows that
\begin{align}
\label{Q-I}
e^{-iT\widetilde{J}}=U e^{-iTJ}U^{\top}=e^{i\phi} U R U^{\top}\equiv e^{i\phi}Q.
\end{align}
Note that the action of $Q$ on $\ket{0}$ is prescribed by \eqref{FR-2}. This similarity transformation is easily found and can be presented as follows. For convenience write $U$ in the form
\begin{align}
U=VR.
\end{align}
Let $V$ be the $(N+1)\times (N+1)$ matrix defined as follows. For $N$ odd, take
\begin{align}
\label{V-Odd}
 V=
\begin{pmatrix}
 \sin \theta &&&&&\cos \theta \\
 &\ddots &&&\udots & \\
 &&\sin \theta& \cos\theta&&\\
 &&\cos \theta& -\sin\theta &&\\
 &\udots &&&\ddots&\\
 \cos \theta &&&&&-\sin \theta
\end{pmatrix},
\end{align}
and for $N$ even, let
\begin{align}
\label{V-Even}
V=
\begin{pmatrix}
\sin \theta &&&&&&\cos\theta\\
& \ddots &&&&\udots&\\
&&\sin\theta&0&\cos\theta &&\\
&& 0 & 1 &0 &&\\
&&\cos \theta & 0 & -\sin\theta &&\\
&\udots &&&&\ddots&\\
\cos \theta &&&&&&-\sin\theta
\end{pmatrix}.
\end{align}
It is immediate to check that $V=V^{\top}$ and that $V^2=\mathbb{1}$. It then follows that $UU^{\top}=\mathbb{1}$. Note also that $\det V=-1$. Obviously $V(0)=R$. The matrix $Q$ introduced in \eqref{Q-I} is thus given by
\begin{align}
VRV=Q,
\end{align}
and is obtained from $V$ by substituting $\theta$ by $2\theta$ in \eqref{V-Odd} for $N$ odd and in \eqref{V-Even} for $N$ even. Obviously $Q^2=\mathbb{1}$. Recall that the PST matrix $J$ is persymmetric $RJR=J$. It is then easy to see that for
\begin{align}
\label{Conju}
\widetilde{J}=U J U^{\top}=VJV,
\end{align}
condition \eqref{FR-2} is satisfied. In fact, not only is this realized but in view of \eqref{Q-I} and the expression for $Q$, we shall have fractional revival between the mirror-symmetric sites $\ell$ and $N-\ell$ since we have
\begin{equation}
e^{-iT\widetilde{J}}=\left\{
\begin{matrix}
& \text{$N$ odd} & \text{$N$ even} \\[.2cm]
\sin 2\theta\,\ket{\ell}+\cos 2\theta\,\ket{N-\ell} & \ell\leq \frac{N-1}{2} & \ell<\frac{N}{2} \\[.1cm]
-\sin 2\theta\,\ket{\ell}+\cos 2\theta\,\ket{N-\ell} & \ell\geq \frac{N+1}{2} & \ell>\frac{N}{2}
\end{matrix}\right.,
\end{equation}
and for $N$ even 
\begin{align}
e^{-i T\widetilde{J}}\,\ket{\textstyle{\frac{N}{2}}}=e^{i\phi}\ket{\textstyle{\frac{N}{2}}}.
\end{align}
To sum up, we have seen that a chain with FR can be obtained from any chain with PST by conjugating the Jacobi matrix of the latter according to \eqref{Conju}. The resulting operator $\widetilde{J}$ is not mirror-symmetric but is seen to be invariant under the one-parameter involution $Q$, that is
\begin{align}
Q\widetilde{J}Q=\widetilde{J}.
\end{align}
It is remarkable that the only modifications or perturbations in the couplings and magnetic fields that arise when passing from $J$ to $\widetilde{J}$ occur in the middle of the chain. Indeed, upon performing the conjugation \eqref{Conju} with \eqref{V-Odd} or \eqref{V-Even}, recalling that $J$ is persymmetric, one finds that the only entries of $\widetilde{J}$ that differ from those of $J$ are
\begin{subequations}
\label{Perturbations}
\begin{align}
\begin{aligned}
\widetilde{J}_{\frac{N+1}{2}}&=J_{\frac{N+1}{2}}\,\cos 2\theta,
\\
\widetilde{B}_{\frac{N\mp 1}{2}}&= B_{\frac{N-1}{2}}\pm J_{\frac{N+1}{2}}\sin 2\theta,
\end{aligned}
\end{align}
for $N$ odd and
\begin{align}
\begin{aligned}
\widetilde{J}_{\frac{N}{2}}&= J_{\frac{N}{2}}(\cos \theta+\sin \theta),
\\
\widetilde{J}_{\frac{N}{2}+1}&=J_{\frac{N}{2}}(\cos \theta-\sin \theta),
\end{aligned}
\end{align}
\end{subequations}
 for $N$ even.
When $N$ is even, only the couplings between the three middle neighbors are altered. When $N$ is odd, it is only the coupling between the two middle neighbors that is affected together with the magnetic field strengths at those two middle sites. Note that if all the $B_{\ell}$ of $J$ are initially zero, $\widetilde{J}$ will only have two Zeeman terms of equal magnitude and opposite sign at $\ell=\frac{N-1}{2}$ and $\ell=\frac{N+1}{2}$. The fact that in these models so few couplings or field strengths of the PST chain need to be adjusted to obtain the chain with FR could prove to be a practical advantage. One can imagine that the calibration would first be done by engineering the couplings so as to reproduce the PST mirror inversion and that thereafter the transformation to the FR mode would not be technically too prohibitive. It is also interesting to remark that it is possible to have no (zero) coupling between two equal parts of the chain and hence two separate chains in fact, and yet to keep some transport. Indeed, when $\theta=\pm \frac{\pi}{4}$ for $N$ even we have $\widetilde{J}_{\frac{N+1}{2}}=0$, $\widetilde{B}_{\frac{N\mp 1}{2}}= B_{\frac{N-1}{2}}\pm J_{\frac{N+1}{2}}$ and when $N$ is odd, $\widetilde{J}_{\frac{N}{2}}=0$, $\widetilde{J}_{\frac{N}{2}+1}=\sqrt{2} J_{\frac{N}{2}}$. It is clear that analytic spin chain models with fractional revival can be obtained from the analytic models with PST that are known by performing the isospectral deformations that we have described in this section. Take again for example the system associated to the Krawtchouk polynomials. Starting with the couplings $J_{\ell}$ and magnetic field $B_{\ell}$ given in \eqref{Recu-Krawtchouk} and modifying them according to  \eqref{Perturbations} will yield a rather simple Hamiltonian $\widetilde{H}$ (with one-excitation sector $\widetilde{J}$) with fractional revival. The exact solvability properties of the perturbed model will be inherited from those of the Krawtchouk chains. For instance, the general transition amplitude between the one-excitation states $\ket{\ell}$ and $\ket{k}$ during time $t$ under the evolution governed by $\widetilde{J}$, that is $\Braket{k}{e^{-i t \widetilde{J}}}{\ell}$ can be obtained directly from the corresponding quantity associated to $J$ and given in \eqref{Kraw-Amp}. Indeed,
\begin{align}
\Braket{k}{e^{-it\widetilde{J}}}{\ell}&=\Braket{k}{V e^{-iTJ} V}{\ell}
=\sum_{m n}V_{mk}V_{n\ell}\,\Braket{m}{e^{-itJ}}{n},
\end{align}
which will yield a sum of (at most) four terms owing to the special form of $V$. Let us mention that some of the couplings \eqref{Perturbations} have appeared in studies of entanglement generation. The cas $\theta=\pi/8$ in \cite{2010_Dai&Feng&Kwek_JPhysA_43_035302} and the case $N$ even in \cite{2010_Kay_IntJQtmInf_8_641}.

To complete the discussion, we shall conclude this section by providing information on the relation that the orthogonal polynomials associated to the Jacobi matrix $\widetilde{J}=V J V$ with fractional revival bear with those attached to the matrix $J$ with PST. This will offer consistency checks and will be of relevance when considering the generic situation when the relative phase $\psi$ of \eqref{Cond-Eigenvalues-2} is arbitrary. Let $\ket{\widetilde{\lambda}_{s}}$ be the eigenstates of $\widetilde{J}$
\begin{align}
\widetilde{J}\,\ket{\widetilde{\lambda}_{s}}=\lambda_{s} \ket{\widetilde{\lambda}_{s}}.
\end{align}
Recall that $\widetilde{J}$ and $J$ have the same spectrum. We have an expansion analogous to \eqref{Expansion} in terms of a different set of orthogonal polynomials $\widetilde{\chi}_{\ell}(\lambda)$:
\begin{align}
\label{Expansion-3}
\ket{\widetilde{\lambda}_{s}}=\sum_{\ell=0}^{N}\sqrt{\widetilde{w}_{s}}\,\widetilde{\chi}_{\ell}(\lambda_{s})\ket{\ell},
\end{align}
where the weights $\widetilde{w}_{s}$ are given by the formula \eqref{W-1} that now reads
\begin{align}
\widetilde{w}_{s}=\frac{\widetilde{h}_{N}}{\widetilde{P}_N(\lambda_{s}) \widetilde{P}_{N+1}'(\lambda_{s})},
\end{align}
with the monic polynomials $\widetilde{P}_{\ell}$ defined by
\begin{align}
\widetilde{P}_{\ell}=\sqrt{\widetilde{h}_{\ell}}\,\widetilde{\chi}_{\ell},\qquad \sqrt{\widetilde{h}_{\ell}}=\widetilde{J}_1\widetilde{J}_2\cdots \widetilde{J}_{\ell}.
\end{align}
From \eqref{Perturbations} we see that
\begin{align}
\sqrt{\frac{\widetilde{h}_{N}}{h_{N}}}=\cos 2\theta.
\end{align}
Now, in view of \eqref{Cond-Gamma-2} and \eqref{Weight-Explicit}, we find that the weights $\widetilde{w}_{s}$ and $w_{s}$ are related as follows:
\begin{align}
\widetilde{w}_{2s}=
\begin{cases}
\gamma \cos 2\theta\, w_{2s} & \text{$N$ odd}
\\
\frac{1}{\gamma} \cos 2\theta\, w_{2s} & \text{$N$ even}
\end{cases},
\quad 
\widetilde{w}_{2s+1}=
\begin{cases}
\frac{1}{\gamma}\cos 2\theta\, w_{2s+1} & \text{$N$ odd}
\\
\gamma \cos 2\theta\, w_{2s+1} & \text{$N$ even}
\end{cases}.
\end{align}
Since $\gamma+\gamma^{-1}=2 \sec 2\theta $ as is readily observed from \eqref{abv-2}, one checks in particular that $\sum_{s}\widetilde{w}_{s}=1$ using  \eqref{Gen-Prop}. With $\ket{\widetilde{\lambda}_{s}}=V\ket{\lambda_{s}}$, using \eqref{Expansion},  one also has
\begin{align}
\label{Expansion-4}
\ket{\widetilde{\lambda}_{s}}=V\ket{\lambda_{s}}=\sum_{\ell,k=0}^{N}\sqrt{w_{s}}\,\chi_{k}(\lambda_{s}) V_{\ell k}\ket{\ell},
\end{align}
and upon comparing \eqref{Expansion-3} with \eqref{Expansion-4}, one finds the relation
\begin{align}
\sqrt{\widetilde{w}_{s}}\,\widetilde{\chi}_{\ell}(\lambda_{s})=\sum_{k=0}^{N}\sqrt{w_{s}}\, V_{\ell k}\,\chi_{k}(\lambda_{s}).
\end{align}
Given the form of $V$ and the property \eqref{Chi-Prop} of the polynomials, one can write 
\begin{align}
\label{Reee-1}
\sqrt{\widetilde{w}_{s}}\,\widetilde{\chi}_{\ell}(\lambda_{s})=\sqrt{w_{s}}\,(V_{\ell, \ell}+(-1)^{N+s}V_{\ell, N-\ell})\,\chi_{\ell}(\lambda_{s}),
\end{align}
together with
\begin{align}
\label{Reee-2}
\sqrt{\widetilde{w}_{s}}\,\widetilde{\chi}_{\frac{N}{2}}(\lambda_{s})=\sqrt{w_{s}}\,\chi_{\frac{N}{2}}(\lambda_{s}),
\end{align}
for $N$ even. In this last instance, note that \eqref{Chi-Prop} gives
\begin{align}
\chi_{\frac{N}{2}}(\lambda_{s})=(-1)^{s}\chi_{\frac{N}{2}}(\lambda_{s}),
\end{align}
meaning that $\chi_{\frac{N}{2}}(\lambda)$ is zero on all the odd eigenvalues: $\chi_{\frac{N}{2}}(\lambda_{2s+1})=0$. The simple trigonometric identities
\begin{align}
\label{75}
\gamma \pm \gamma^{-1}=\tan \left(\frac{\pi}{4}-\theta\right)\pm \cot \left(\frac{\pi}{4}-\theta\right)=
\begin{cases}
2 \sec 2\theta &
\\
-2 \tan 2\theta &
\end{cases}
\end{align}
allow to show that 
\begin{align}
\label{76}
\gamma \cos 2\theta= (\sin \theta -\cos \theta)^2
\qquad
\gamma^{-1}\cos 2\theta=(\sin \theta +\cos \theta)^2
\end{align}
Using these relations and examining each case, one checks that \eqref{Reee-1} and \eqref{Reee-2} imply
\begin{subequations}
\label{Evaluations}
\begin{align}
\begin{aligned}
\widetilde{\chi}_{\ell}(\lambda_{s})&=\chi_{\ell}(\lambda_{s}),\qquad \ell \leq \frac{N-1}{2},
\\
\widetilde{\chi}_{\ell}(\lambda_{2s})&=\gamma^{-1}\chi_{\ell}(\lambda_{2s}),\qquad \ell \geq \frac{N+1}{2},
\\
\widetilde{\chi}_{\ell}(\lambda_{2s+1})&=\gamma \chi_{\ell}(\lambda_{2s+1}),\qquad \ell \geq \frac{N+1}{2},
\end{aligned}
\end{align}
for $N$ odd and
\begin{align}
\begin{aligned}
\widetilde{\chi}_{\ell}(\lambda_{s})&=\chi_{\ell}(\lambda_{s}),\qquad \ell <\frac{N}{2},
\\
\widetilde{\chi}_{\frac{N}{2}}(\lambda_{2s})&=\frac{\chi_{\frac{N}{2}}(\lambda_{2s})}{\sin\theta+\cos \theta},\qquad \ell=\frac{N}{2},
\\
\widetilde{\chi}_{\frac{N}{2}}(\lambda_{2s+1})&=\chi_{\frac{N}{2}}(\lambda_{2s+1})=0,\qquad \ell=\frac{N}{2},
\\
\widetilde{\chi}_{\ell}(\lambda_{2s})&=\gamma \chi_{\ell}(\lambda_{2s}),\qquad \ell >\frac{N}{2},
\\
\widetilde{\chi}_{\ell}(\lambda_{2s+1})&=\gamma^{-1}\chi_{\ell}(\lambda_{2s+1}),\qquad \ell>\frac{N}{2},
\end{aligned}
\end{align}
\end{subequations}
for $N$ even. The proper sign should be chosen in taking square roots of the relations \eqref{76} so that \eqref{Cond-Gamma-2} is fulfilled.

Consider now the difference between the monic polynomials $\widetilde{P}_{N}$ and $P_{N}$:
\begin{align}
\widetilde{P}_{N}-P_{N}=\sqrt{h_{N}}(\cos 2\theta \,\widetilde{\chi}_{N}-\chi_N).
\end{align}
This is a polynomial of degree $N-1$. Evaluating on the spectral points we have
\begin{align}
\widetilde{P}_{N}(\lambda_{2s})-P_{N}(\lambda_{2s})&=\sqrt{h_{N}}\,(1-\gamma^{-1}\cos 2\theta),
\\
\widetilde{P}_{N}(\lambda_{2s+1})-P_{N}(\lambda_{2s+1})&=\sqrt{h_{N}}\,(\gamma \cos 2\theta -1),
\end{align}
for $N$ odd. The right-hand sides are interchanged for $N$ even. It is readily checked with \eqref{75} that 
\begin{align}
(1-\gamma^{-1}\cos 2\theta)=\gamma \cos 2\theta-1.
\end{align}
Hence for $N$ odd and even,
\begin{align}
\widetilde{P}_{N}(\lambda_{s})-P_{N}(\lambda_{s})=\sqrt{h_{N}}(\gamma \cos 2\theta -1).
\end{align}
This shows that $\widetilde{P}_{N}-P_{N}$, a polynomial of degree $N-1$, is equal to a constant for $N$ points and must hence be identically equal to that constant. We thus have
\begin{align}
\label{ddp}
\widetilde{P}_{N}(\lambda_{s})=P_{N}(\lambda_{s})+\zeta_0,
\end{align}
with
\begin{align}
\zeta_0=J_1J_2\cdots J_{N}(\gamma \cos 2\theta-1).
\end{align}
From the knowledge of $\widetilde{P}_{N+1}(\lambda_{s})=P_{N+1}(\lambda_{s})$ and of $\widetilde{P}_{N}(\lambda_{s})$, using the recurrence relations, it is possible to show by induction that
\begin{subequations}
\label{84}
\begin{align}
\widetilde{P}_{\ell}=P_{\ell},
\end{align}
and 
\begin{align}
\label{BB}
\widetilde{P}_{N-\ell}=P_{N-\ell}+\zeta_{\ell} P_{\ell},
\end{align}
for
\end{subequations}
\begin{align}
\ell=
\begin{cases}
0, \ldots, \frac{N-1}{2} & \text{$N$ odd}
\\
0,\ldots, \frac{N}{2}-1 & \text{$N$ even}
\end{cases},
\end{align}
with in addition 
\begin{align}
\widetilde{P}_{\frac{N}{2}}(\lambda)=P_{\frac{N}{2}}(\lambda)=(\lambda-\lambda_1)(\lambda-\lambda_3)\cdots (\lambda-\lambda_{N-1}),
\end{align}
when $N$ is even. The constant $\zeta_{\ell}$ in \eqref{BB} is given by
\begin{align}
\zeta_{\ell}=\frac{\zeta_0}{J^2_{N+1-\ell}\cdots J_{N-1}^2 J_{N}^2}.
\end{align}
Details will be given elsewhere \cite{Tsujimoto_2015}. The fact that the polynomials $\widetilde{P}_{\ell}$ are equal to the unperturbed polynomials $P_{\ell}$ for the first half of the indices/degrees was expected because the recurrence coefficients are the same up that point. It is readily checked that the evaluations \eqref{Evaluations} on the spectral points are entirely consistent with the formulas \eqref{84} when one allows for  \eqref{14}.
\section{The bi-lattices models and para-Krawtchouk polynomials}
We saw in Section 4 that within the class of $XX$ spin chains with non-uniform nearest neighbor couplings, the general conditions in order to have fractional revival at two sites are two-fold. One, the spectrum $\{\lambda_{s}\}$ of the one-excitation Hamiltonian $J$ must be comprised of the points of the bi-lattice \eqref{Bi-Lattice} or of an ordered subset of those grid points resulting from the removal of consecutive eigenvalues. Second, the transition matrix that diagonalizes $J$ must be made out of polynomials that are orthogonal with respect to the weight $w_{s}$ given by
\begin{align}
\label{weights-2}
\begin{aligned}
w_{2s}=
\begin{cases}
-\frac{\gamma \sqrt{h_{N}}}{P_{N+1}'(\lambda_{2s})} & \text{$N$ odd}
\\
\frac{\sqrt{h_{N}}}{\gamma P_{N+1}'(\lambda_{2s})} & \text{$N$ even}
\end{cases},
\qquad 
w_{2s+1}=
\begin{cases}
\frac{\sqrt{h_{N}}}{\gamma P_{N+1}'(\lambda_{2s+1})} & \text{$N$ odd}
\\
-\frac{\gamma \sqrt{h_{N}}}{P_{N+1}'(\lambda_{2s+1})} & \text{$N$ even}
\end{cases},
\end{aligned}
\end{align}
with $\gamma$ the positive root of \eqref{Cond-Gamma} and $P_{N+1}'(\lambda)$ as before, the derivative of the characteristic polynomial. Finding the specifications of the corresponding spin chain amounts to an inverse spectral problem that can be solved by finding the polynomials occurring in the transition matrix and thereafter their recurrence coefficients which are the entries of $J$. In Section 4 still, we pointed out that there is an interesting special case that arises when the relative phase $\psi=\frac{\pi}{2}$. When this is so, $\gamma=1$, $w_{s}$ is given by \eqref{Weight-Explicit} and we know that $J$ is mirror-symmetric. This case has been considered in \cite{2015_Banchi&Compagno&Bose_PhysRevA_91_052323} and we shall discuss it in detail here. The authors of \cite{2015_Banchi&Compagno&Bose_PhysRevA_91_052323} have determined numerically the persymmetric Jacobi matrix in the perfectly balanced situation $\theta=\pi/8$. We shall indicate that there is in fact an exact description for any $\theta$. Indeed, chains with bi-lattice spectra and mirror-symmetric couplings have been analyzed with the help of the para-Krawtchouk polynomials that two of us have identified and characterized in \cite{2012_Vinet&Zhedanov_JPhysA_45_265304}. Since their Jacobi matrix is persymmetric, these models are poised to admit PST. The circumstances under which they shall exhibit PST in addition to FR will be discussed. We shall conclude the section by returning to the general case. We shall explain that it can be realized by combining the construction of the persymmetric matrices associated to bi-lattices and para-Krawtchouk polynomials with the isospectral deformations described in Section 5 and possibly surgeries.

When $\psi=\pi/2$, the spectral condition \eqref{Cond-Eigenvalues-2} becomes
\begin{align}
e^{-iT\lambda_{s}}=e^{i\phi}\left[\sin 2\theta +(-1)^{N+s}\,i\,\cos 2\theta \right],
\end{align}
which amounts to
\begin{align}
e^{-iT\lambda_{2s}}=e^{i\phi}e^{i\left(\frac{\pi}{2}-2\theta\right)},\qquad e^{-iT\lambda_{2s+1}}=e^{i\phi}e^{-i\left(\frac{\pi}{2}-2\theta\right)}.
\end{align}
Let us now calculate the transition amplitude $\Braket{k}{e^{-iTJ}}{\ell}$ in analogy with what was done in Section 3 (see \eqref{Calculation}). Assume that $N$ is odd, one has
\begin{align}
\begin{aligned}
&\Braket{k}{e^{-iT J}}{\ell}
\\
&=\sum_{2s} e^{-iT\lambda_{2s}}\,w_{2s}\,\chi_{\ell}(\lambda_{2s})\chi_{k}(\lambda_{2s}) +\sum_{2s+1}e^{-iT\lambda_{2s+1}}\,w_{2s+1}\,\chi_{\ell}(\lambda_{2s+1})\chi_k(\lambda_{2s+1})
\\
&=e^{i\phi}e^{-i\pi/2}\Big[ \cos 2\theta \,\Big(\sum_{2s}w_{2s}\,\chi_{\ell}(\lambda_{2s})\chi_{k}(\lambda_{2s})
-\sum_{2s+1}w_{2s+1}\,\chi_{\ell}(\lambda_{2s+1})\chi_{k}(\lambda_{2s+1})\Big)
\\
&+i \sin 2\theta\,\Big( \sum_{2s}w_{2s}\,\chi_{\ell}(\lambda_{2s})\chi_{k}(\lambda_{2s})+\sum_{2s+1}w_{2s+1}\,\chi_{\ell}(\lambda_{2s+1})\chi_{k}(\lambda_{2s+1})\Big)\Big].
\end{aligned}
\end{align}
Making use of \eqref{Chi-Prop}, one finds that
\begin{align}
\Braket{k}{e^{-iTJ}}{\ell}=e^{i\phi}\left[\delta_{\ell k}\sin 2\theta+i \cos 2\theta \delta_{N-\ell, k}\right],
\end{align}
which shows that
\begin{align}
e^{-iTJ}\ket{\ell}=e^{i\phi}\left[\sin 2\theta \ket{\ell}+i \cos 2\theta \ket{N-\ell}\right],\qquad \ell=0,1,\ldots, N.
\end{align}
As observed in \cite{2015_Banchi&Compagno&Bose_PhysRevA_91_052323}, we see that a state localized at site $\ell$ will be revived at the sites $\ell$ and $N-\ell$. In matrix form, we have found that 
\begin{align}
\label{93}
e^{-iTJ}=e^{i\phi}
\begin{pmatrix}
\sin 2\theta & & & i \cos 2\theta
\\
&\ddots & \udots &
\\
& \udots & \ddots &
\\
i \cos 2\theta &&& \sin 2\theta
\end{pmatrix},
\end{align}
for $N$ odd. In the same way, one shows that for $N$ even
\begin{align}
\label{94}
e^{-iTJ}=e^{i\phi}
\begin{pmatrix}
\sin 2\theta &&&& i \cos 2\theta
\\
& \ddots &&\udots &
\\
&&e^{i\left(\frac{\pi}{2}-2\theta \right)} &&
\\
&\udots && \ddots &
\\
i \cos 2\theta &&&& \sin 2\theta
\end{pmatrix}.
\end{align}
The set of couplings and magnetic field strengths for which Hamiltonians of the form \eqref{H} will lead to this behavior has been provided explicitly in \cite{2012_Vinet&Zhedanov_JPhysA_45_265304}. They happen to be formed of the recurrence coefficients of the orthogonal polynomials that have been called the para-Krawtchouk polynomials. These OPs are precisely those that are associated to persymmetric Jacobi matrices with the bi-lattice spectra
\begin{align}
\label{latt}
\overline{x}_{s}=s+\frac{1}{2}(\delta-1)(1-(-1)^{s}).
\end{align}
They have been constructed with the help of the Euclidean algorithm, described in \cite{2012_Vinet&Zhedanov_PhysRevA_85_012323}, from the knowledge of the two polynomials $\overline{P}_{N+1}$ and $\overline{P}_{N}$, the former being prescribed by the spectrum and the latter by the mirror symmetry. They have been named para-Krawtchouk polynomials on the one hand because their spectrum coincides, when $N\rightarrow \infty$, with that of the parabosonic oscillator \cite{1994_Rosenblum} and on the other hand because they become the standard Krawtchouk polynomials when $\delta=1$. Their recurrence coefficients and properties are given in \cite{2012_Vinet&Zhedanov_JPhysA_45_265304}. One has for $N$ odd
\begin{align}
\label{96}
\overline{B}_{\ell}=\frac{N-1+\delta}{2},\qquad \overline{J}_{\ell}=\frac{1}{2}\sqrt{\frac{\ell(N+1-\ell)((N+1-2\ell)^2-\delta^2)}{(N-2\ell)(N-2\ell+2)}},
\end{align}
and for $N$ even 
\begin{align}
\label{97}
\begin{aligned}
\overline{B}_{\ell}&=\frac{N-1+\delta}{2}+\frac{(\delta-1)(N+1)}{4}\left(\frac{1}{2\ell-N-1}-\frac{1}{2\ell+1-N}\right),
\\
\overline{J}_{\ell}&=\frac{1}{2}\sqrt{\frac{\ell(N+1-\ell)((2\ell-N-1)^2-(\delta-1)^2)}{(2\ell-N-1)^2}},
\end{aligned}
\end{align}
for $\ell=0,1,\ldots, N$. These formulas correspond to the lattice \eqref{latt}. It is easy to see from the recurrence relation
\begin{align}
\overline{x}\,\overline{P}_{\ell}(\overline{x})=\overline{P}_{\ell+1}(\overline{x})+\overline{B}_{\ell} \overline{P}_{\ell}(\overline{x})+\overline{J}_{\ell}^2 \overline{P}_{\ell-1}(\overline{x}),
\end{align}
of the monic polynomials for instance, that an affine transformation of the lattice points
\begin{align}
x_{s}=a \overline{x}_{s}+b,
\end{align}
will lead to orthogonal polynomials $P_{\ell}(x)$ with recurrence coefficients given by
\begin{align}
\label{100}
B_{\ell}=a \overline{B}_{\ell}+b,\qquad J_{\ell}=a \overline{J}_{\ell}.
\end{align}
Note that the diagonal terms $\overline{B}_{\ell}$, that is the magnetic fields, are the same at every site for $N$ odd; see \eqref{96}. They can thus be made equal to zero by an affine transformation. This is not so for $N$ even however. Comparing \eqref{Bi-Lattice} with \eqref{latt} and making use of \eqref{100}, we see that by choosing the global phase $\phi$ to be 
\begin{align}
\phi=\frac{\pi}{2}(N-1+\delta)=
\begin{cases}
\frac{\pi(N+1)}{2} & \text{$N$ odd}
\\
\frac{\pi(N-1)}{2}& \text{$N$ even}
\end{cases},
\end{align}
and in view of \eqref{mu}-\eqref{delta}, the spin chains with the fractional revival features described in this section have their couplings and magnetic fields given by
\begin{align}
\label{102}
B_{\ell}=0,\qquad J_{\ell}=\frac{\pi}{T} \overline{J}_{\ell},
\end{align}
for $N$ odd and 
\begin{align}
\label{103}
B_{\ell}=-\frac{\theta}{T}(N+1)\left(\frac{1}{2\ell-N-1}-\frac{1}{2\ell+1-N}\right),\quad J_{\ell}=\frac{\pi}{T} \overline{J}_{\ell},
\end{align}
for $N$ even, with $\overline{J}_{\ell}$ given by \eqref{96} and \eqref{97} where $\delta=1+4\theta/\pi$ for $N$ odd and $\delta=1-4\theta/\pi$ for $N$ even. As observed also numerically in \cite{2015_Banchi&Compagno&Bose_PhysRevA_91_052323}, relative to the coefficients of the Krawtchouk chain given in \eqref{Recu-Krawtchouk}, the magnetic fields remains zero for $N$ odd while they are proportional to $\theta$ for $N$ even. Note that the Krawtchouk chain parameters are recovered when $\theta=0$ and that contrary to the isospectral models with fractional revival covered in the last section, here, all the $J_{\ell}$ are modified in comparison with those of \eqref{Recu-Krawtchouk}. Interestingly, the para-Krawtchouk models have been shown in \cite{2012_Vinet&Zhedanov_JPhysA_45_265304} to enact PST for
\begin{align}
\delta=\frac{M_1}{M_2},
\end{align}
where $M_1$ and $M_2$ are positive co-prime integers and $M_1$ is odd. Let us here explain how spin chains that lead to fractional revival can also exhibit perfect state transfer. To that end, introduce the Hadamard matrices
\begin{subequations}
\label{Hadamard}
\begin{align}
H=\frac{1}{\sqrt{2}}
\begin{pmatrix}
1&&&&&1
\\
&\ddots &&&\udots &
\\
&&1&1&&
\\
&&1&-1&&
\\
&\udots &&&\ddots &
\\
1&&&&&-1
\end{pmatrix},
\end{align}
for $N$ odd and 
\begin{align}
H=\frac{1}{\sqrt{2}}
\begin{pmatrix}
1&&&&&&1
\\
&\ddots &&&&\udots&
\\
&&1&&1&&
\\
&&&\sqrt{2}&&&
\\
&&1&&-1&&
\\
&\udots &&&&\ddots&
\\
1&&&&&&-1
\end{pmatrix},
\end{align}
\end{subequations}
for $N$ even. It is readily seen that 
\begin{align}
e^{-iTJ}=H\,U(\alpha)\,H,
\end{align}
with $\alpha=\frac{\pi}{2}-2\theta$ and $U(\alpha)$ the unitary diagonal matrix with elements
\begin{align}
U_{ij}(\alpha)=\delta_{ij}
\begin{cases}
e^{i\alpha} & i,j=0,\ldots, \lfloor \frac{N}{2}\rfloor
\\
e^{-i\alpha} & i,j=\lfloor \frac{N}{2}+1\rfloor, \ldots, N
\end{cases},
\end{align}
where $\lfloor x\rfloor$ is the integer part of $x$. For $M$ an integer, it thus follows that
\begin{align}
e^{-iMTJ}=H\,U(M\alpha)\,H,
\end{align}
since $H^2=1$. Therefore, after a time $MT$ one has $\frac{\pi}{2}-2\theta\rightarrow M\left(\frac{\pi}{2}-2\theta\right)$. Express now the manifestation of fractional revival in the form
\begin{align}
e^{-iTJ}\ket{0}=e^{i\phi}\left[\cos\left(\frac{\pi}{2}-2\theta\right)\ket{0}+i \sin \left(\frac{\pi}{2}-2\theta\right)\ket{N}\right].
\end{align}
It follows that 
\begin{align}
e^{-iM T J}\ket{0}=e^{i\phi}\left[\cos M\left(\frac{\pi}{2}-2\theta\right)\ket{0}+i \sin M \left(\frac{\pi}{2}-2\theta\right)\ket{N}\right].
\end{align}
Perfect state transfer will occur if 
\begin{align}
\label{Cnd-3}
M\left(\frac{\pi}{2}-2\theta\right)=M_1\left(\frac{\pi}{2}\right),
\end{align}
with $M_1$ an arbitrary odd number since then $e^{iMTJ}\ket{0}=e^{i\widetilde{\phi}}\ket{N}$ with $\widetilde{\phi}$ some phase factor. Condition \eqref{Cnd-3} is readily seen to be equivalent to \eqref{Hadamard}; when $N$ is even and $\delta=1-\frac{4\theta}{\pi}$ it is immediate and when $N$ is odd and $\delta=1+\frac{4\theta}{\pi}$ one uses the properties of the cosine to conclude. Hence when \eqref{Cnd-3} is verified these spin chains with FR will also exhibit PST at time $MT$. Take for example the perfectly balanced case of FR which occurs when $\theta=\pi/8$, one has then $\delta=1+\frac{4\theta}{\pi}=\frac{3}{2}$ for $N$ odd or $\delta=1-\frac{4\theta}{\pi}=\frac{1}{2}$ for $N$ even and it follows that PST will also happen. The evolution will go like this: at time $t=T$, the packet initially at site $0$ is revived at $0$ and $N$, at $t=2T$ it is perfectly transferred at $N$, at $t=3T$ is is revived again at $0$ and $N$, at $t=4T$ it perfectly returns to $0$, and so on. 

Let us now complete our systematic analysis by considering the general case where the phase $\psi$ is arbitrary. As stated at the beginning of the section the polynomials that will determine the general Hamiltonians are orthogonal with respect to the weights \eqref{weights-2} associated to the bi-lattices \eqref{Bi-Lattice}. We now understand that we can obtain these polynomials in two steps. First, we determine the para-Krawtchouk polynomials associated to a bi-lattice with
\begin{align}
\label{abv-4}
\delta=1\pm \frac{4\sigma}{\pi},\quad \binom{\text{$N$ odd}}{\text{$N$ even}},
\end{align}
where $\sigma$ is given by 
\begin{align}
\label{sigma-def}
\sigma=\frac{\pi +\eta-\xi}{4}.
\end{align}
The Jacobi matrix $J$ is then given by \eqref{102} and \eqref{103} with $\theta$ replaced by $\sigma$ and again using the $+$ sign in \eqref{abv-4} when $N$ is odd and the $-$ sign when $N$ is even. At this point we have that $e^{-iTJ}$ is given by  
\eqref{93} or \eqref{94} with $\theta$ again replaced by $\sigma$ and $\phi$ by
\begin{align}
\label{phibar}
\overline{\phi}=\frac{\pi}{4}(N\pm 1)+\frac{1}{2}(\eta+\xi),\quad \binom{\text{$N$ odd}}{\text{$N$ even}}.
\end{align}
This gives us the polynomials $P_{\ell}$ that are associated to the bi-lattice \eqref{Bi-Lattice} but are orthogonal with respect to the weights \eqref{weights-2} with $\gamma=1$ (corresponding to a persymmetric Jacobi matrix) and the $h_{N}$ of the para-Krawtchouk polynomials. The required polynomials that are properly orthogonal against the weights \eqref{weights-2} are the perturbed polynomials $\widetilde{P}_{\ell}$ defined in \eqref{84} with
\begin{align}
\zeta_0=J_1J_2\cdots J_{N}(\widetilde{\gamma}\cos 2\tau-1),
\end{align}
with $\tau$ a new angle so that $\widetilde{\gamma}-\widetilde{\gamma}^{-1}=-2\tan 2\tau$ and $J_{\ell}$ the recurrence coefficients determined in the first step. These polynomials $\widetilde{P}_{\ell}$ will yield through their recurrence relation, the parameters $\widetilde{J}_{\ell}$ and $\widetilde{B}_{\ell}$ of the generic chain. The modifications relative to the para-Krawtchouk coefficients are given by the formulas \eqref{Perturbations} with $\theta$ replaced by $\tau$. The determination of $e^{-iT\widetilde{J}}$ is achieved by conjugating $e^{-iTJ}$ as given in \eqref{93} or \eqref{94} respectively with the matrix $V$ of \eqref{V-Even} or \eqref{V-Odd} with $\theta$ replaced by $\tau$. Thus are determined the Hamiltonians (within the class considered) that have general fractional revivals at two sites. Note that as needed, the two-step process has introduced two angles $\sigma$ and $\tau$. The correspondence with the original parameters $\theta$ and $\psi$ can be obtained by determining explicitly $e^{-iT\widetilde{J}}$ as indicated before and identifying the coefficients so that $e^{-iT\widetilde{J}}\ket{0}=e^{i\phi}\left[\sin 2\theta \ket{0}+i e^{i\psi}\cos 2\theta \ket{N}\right]$. This leads to the following relations
\begin{subequations}
\label{cccc}
\begin{align}
\label{aaaa}
e^{i\overline{\phi}}(\sin 2\sigma+i \cos 2\sigma \sin 2\tau)=e^{i\phi}\sin 2\theta,
\\
\label{bbbb}
i e^{i\overline{\phi}}\cos 2\sigma \cos 2\tau=e^{i(\phi+\psi)} \cos 2\theta,
\end{align}
\end{subequations}
with $\overline{\phi}$ given by \eqref{phibar}. These conditions are the same for $N$ odd or even provided the appropriate $\overline{\phi}$ is chosen. Equation \eqref{bbbb} immediately leads to 
\begin{align}
\phi=\overline{\phi}-\psi+\frac{\pi}{2}+2n\pi,\qquad n\in \mathbb{Z}.
\end{align}
The real part of \eqref{aaaa} yields
\begin{align}
\label{abv-6}
\sin 2\sigma=\sin 2\theta \sin \psi,
\end{align}
which must be identically satisfied. Upon writing the above equation in the form
\begin{align}
\cos \left(\frac{\xi-\eta}{2}\right)=\sin 2\theta \sin \psi,
\end{align}
using \eqref{sigma-def}, that \eqref{abv-6} holds is verified from trigonometric identities having recalled the definitions of $\xi$ and $\eta$ (see the sentence after \eqref{abcd}) and observed that $\sin \xi$ and $\sin \eta$ must have opposite signs. There then remains from \eqref{cccc} the conditions
\begin{align}
\cos 2\tau&=\cos 2\theta \mathrm{cosec}\,\left(\frac{\xi-\eta}{2}\right),
\\
\sin  2\tau&=\sin 2\theta \cos \psi \mathrm{cosec}\,\left(\frac{\xi-\eta}{2}\right),
\end{align}
which determine $\tau$.
\section{Conclusion}
Let us summarize our findings. We have completely characterized the $XX$ spin chains with nearest neighbor couplings that admit fractional revival at two sites. There are two basic ways according to which FR can be realized. One is via isospectral deformations of chains with the PST property and the other is by a mirror-symmetric set of couplings corresponding to the recurrence coefficients of the para-Krawtchouk polynomials. Hamiltonians with FR at two sites controlled by two arbitrary parameters are obtained by compounding these two approaches. The second approach comes with a complete set of couplings and magnetic fields while the first approach only sees the modification of a few central coefficients of a parent PST chain. The time $T$ for FR occurrence doest not depend on the length of the chain. The first method can be applied to any PST chain to obtain a chain with FR. Assuming the model is exactly solvable to start with, it will remain so under the isospectral deformation. The models corresponding to the second way are analytic and may exhibit PST in addition to FR. A note is in order here. In principle all chains with FR at two sites can be obtained from the generic two-parameter models by surgeries. Indeed, since we are dealing with spectra that are finite, any admissible set of eigenvalues can be obtained by removing levels from a bi-lattice chosen as large as required. With every such removal, the analytic expressions for the chain parameters will become more and more involved thus obscuring the exact solvability property.

It has been indicated that information transfer can be achieved with spin chains showing FR at two sites. Knowing that the clone of the initial information will be at the end of the chain with definite probability $(\cos 2\theta)^2$ at the prescribed time $T$, the end site content at that time can thus be used as input to some quantum process or computation with the effect that the final output of the computation will provide the right answer with known probability related to $(\cos 2\theta)^2$. Observe that this probability can be tuned by setting correspondingly the chain couplings. Note also that the presence of another clone at the site $(\ell=0)$ where the data is entered could be used periodically in an experimental or practical context to check that transmission is proceeding without alterations since the outcomes at $\ell=0$ and at $\ell=N$ are correlated. It has also been pointed out \cite{2010_Kay_IntJQtmInf_8_641, 2015_Banchi&Compagno&Bose_PhysRevA_91_052323, 2010_Dai&Feng&Kwek_JPhysA_43_035302} that balanced perfect revival can generate entanglement. Indeed it is readily observed that for $\theta=\pi/8$ the sites $0,1, N-1$ and $N$ for instance, will support at time $T$ the entangled state $\ket{\uparrow}\ket{\downarrow}+\ket{\downarrow}\ket{\uparrow}$. 

Another question has to do with precision. Throughout this paper we have looked for situations where FR occurs with probability 1. This could be unduly stringent in view of the unavoidable instrumental error for instance. In fact, it would suffice in that perspective to consider situations where FR can happen with probability as close to 1 as desired. This question has been analyzed in the case of full revival and has been referred to as almost perfect state transfer (APST) \cite{2012_Vinet&Zhedanov_PhysRevA_86_052319} or pretty good state transfer \cite{2012_Godsil&Kirkland&Severini&Smith_PhysRevLett_109_050502}. One may assume that the isospectral deformations of a chain with APST will lead to chains with almost perfect fractional revival (APFR). Furthermore it has been shown in \cite{2012_Vinet&Zhedanov_JPhysA_45_265304} that the para-Krawtchouk chains  admit APST for a time $T$ independent of $N$ if the bi-lattice parameter $\delta$ is irrational, they should thus admit APFR in those cases too. The robustness of FR in the para-Krawtchouk model has also been checked numerically in \cite{2015_Banchi&Compagno&Bose_PhysRevA_91_052323}. 

Finally, it would be of great interest to study the possibilities for fractional revival at more than two sites. It is known that any unitary matrix can be presented in a form with 
two diagonals and 2 antidiagonals \cite{1993_Watkins_SIAMRev_35_430}; this is related to CMV theory \cite{2003_Cantero&Moral&Velazquez_LinAlgAppl_362_29}. Assume that $e^{-iTJ}$ is in that form in the register basis $\ket{\ell}$, where $\ell=0,1,\ldots, N$. This implies revival at up to four sites. A relevant question is to determine the Hamiltonians $H$ with their one-excitation restrictions $J$ that will lead to such unitaries. This is likely to involve operators beyond the realm of nearest-neighbor interactions. We hope to report on this question in the near future.
\section*{Acknowledgments}
The authors would like to thank L. Banchi, S. Bose, G. Coutinho, M. Christandl and S. Severini for their collegial input. While this paper was being completed we were informed that L. Banchi and G. Coutinho had obtained in a different way the fractional revival described in Section 4 using the same perturbed polynomials that we have identified. We are very grateful that they shared their results with us prior to publication. VXG holds a scholarship from the Natural Science and Engineering Research Council of Canada (NSERC). The research of LV is supported in part by NSERC. AZ would like to thank the Centre de recherches math\'ematiques for its hospitality.
\section*{References}

\end{document}